\documentclass[twocolumn, fleqn]{doc_class}

\usepackage{lineno}

\usepackage[super, comma, sort&compress]{natbib}

\usepackage{multibib}
\newcites{m}{References}
\bibliographystylem{bib_style}
\newcites{met}{References (continued)}
\bibliographystylemet{bib_style}
\newcites{sup}{References (continued)}
\bibliographystylesup{bib_style}

\usepackage{newtxtext, newtxmath}

\usepackage[T1]{fontenc}
\usepackage[utf8]{inputenc}

\providecommand{\apj}{Astrophysical Journal}
\providecommand{\apjl}{Astrophysical Journal, Letters}
\providecommand{\apjs}{Astrophysical Journal, Supplement}
\providecommand{\aap}{Astronomy and Astrophysics}
\providecommand{\aapr}{The Astronomy and Astrophysics Review}
\providecommand{\mnras}{Monthly Notices of the RAS}

\providecommand{\nat}{Nature}
\providecommand{\araa}{Annual Review of Astronomy and Astrophysics}

\providecommand{\nphysa}{Nuclear Physics A}

\usepackage{ae, aecompl}
\usepackage{graphicx}
\usepackage[dvipsnames, hyperref]{xcolor}
\usepackage{xspace}
\usepackage{soul}
\usepackage{booktabs}
\usepackage{adjustbox}
\usepackage{comment}
\usepackage{float}

\usepackage{amsmath}
\usepackage{commath}
\usepackage{siunitx}
\usepackage{nth}

\usepackage[capitalise, noabbrev]{cleveref}
\usepackage{csvsimple}
\usepackage{multirow}
\usepackage{array}
\newcolumntype{P}[1]{>{\centering\arraybackslash}p{#1}}

\crefname{figure}{Fig.}{Figs.}
\crefname{equation}{equation}{equations}

\usepackage[leftmargin=0pt,rightmargin=0pt,skipabove=2pt,skipbelow=2pt,innerleftmargin=5pt,innerrightmargin=5pt,innertopmargin=5pt,innerbottommargin=5pt,linewidth=0pt,innerlinewidth=0pt,middlelinewidth=0pt,outerlinewidth=0pt,roundcorner=0pt]{mdframed}

\newcommand{\methodslink}{\hyperref[sec:methods]{Methods}}

\hypersetup{
    unicode=true,
    final=true,
    plainpages=false,
    pdfstartview=FitV,
    pdftoolbar=false,
    pdfmenubar=true,
    bookmarksopen=true,
    bookmarksnumbered=true,
    breaklinks=true,
    linktocpage,
    colorlinks=true,
    linkcolor=Blue,
    urlcolor=Blue,
    citecolor=Blue,
    anchorcolor=green
}
\AtBeginDocument{
    \hypersetup{
        pdftitle={Globular cluster abundance patterns inherited from giant molecular clouds},
        pdfauthor={W. McClymont et al.},
        pdfsubject={},
        pdfkeywords={Globular clusters, Chemical abundance patterns, Giant molecular clouds, Galaxy evolution, Cosmological simulations, Interstellar medium}
    }
}

\usepackage{paralist}
\usepackage{enumitem}
\setlist[enumerate]{leftmargin=*,align=left}
\makeatletter
    \def\@biblabel#1{\@ifnotempty{#1}{#1}}
    \def\NAT@anchor#1#2{
        \hfilneg\hyper@natanchorstart{#1\@extra@b@citeb}
        #2.
        \hyper@natanchorend
    }
\makeatother

\renewenvironment{thebibliography}[1]{%
    \newpage
    \section*{\normalsize \refname}
    \setlength{\parindent}{0pt}
    \inparaenum
}{\endinparaenum}

\DeclareRobustCommand{\VAN}[3]{#2}
\let\VANthebibliography\thebibliography
\def\thebibliography{\DeclareRobustCommand{\VAN}[3]{##3}\VANthebibliography}

\title[Globular cluster abundances from giant molecular clouds]{Globular cluster abundance patterns inherited from giant molecular clouds}

\author[]{{William McClymont$^{\hyperlink{inst:Kavli}{1}, \hyperlink{inst:Cav}{2}}$\thanks{E-mail: \href{mailto:wjm50@cam.ac.uk}{wjm50@cam.ac.uk}},
Vasily Belokurov$^{\hyperlink{inst:IoA}{3}}$,
Sandro Tacchella$^{\hyperlink{inst:Kavli}{1}, \hyperlink{inst:Cav}{2}}$,
Rahul Kannan$^{\hyperlink{inst:York}{4}}$,
Aaron Smith$^{\hyperlink{inst:Dallas}{5}}$,}
\newauthor{
Ewald Puchwein$^{\hyperlink{inst:AIP}{6}}$,
Enrico Garaldi$^{\hyperlink{inst:IPMU}{7}}$,
Mark Vogelsberger$^{\hyperlink{inst:MIT}{8}}$,
Stephanie Monty$^{\hyperlink{inst:CIERA}{9}, \hyperlink{inst:NMSU}{10}}$,
Josh Borrow$^{\hyperlink{inst:Penn}{11}}$,}
\newauthor{
Laura Keating$^{\hyperlink{inst:IfA}{12}}$,
Xuejian Shen$^{\hyperlink{inst:MIT}{8}}$,
Zihao Wang$^{\hyperlink{inst:MIT}{8}, \hyperlink{inst:PKU}{13}}$,
Oliver Zier$^{\hyperlink{inst:CfA}{14}}$,
Changhyun Cho$^{\hyperlink{inst:York}{4}}$,
Yuki Isobe$^{\hyperlink{inst:Kavli}{1}, \hyperlink{inst:Cav}{2}}$,}
\newauthor{
Xihan Ji$^{\hyperlink{inst:Kavli}{1}, \hyperlink{inst:Cav}{2}}$,
Roberto Maiolino$^{\hyperlink{inst:Kavli}{1}, \hyperlink{inst:Cav}{2}}$,
and Giulia Pruto$^{\hyperlink{inst:IfA}{12}}$}
}

\pubyear{2025}

\begin{document}
\maketitle

\begin{abstract}
    \begin{mdframed}[backgroundcolor=black!5]
        Globular clusters exhibit large star-to-star variations and anticorrelations in their light element abundances that are commonly interpreted in terms of in-cluster self-enrichment, in which ejecta from early-forming cluster stars pollute the gas from which later stars form over millions of years. Yet proposed self-enrichment scenarios suffer from a severe mass-budget problem or invoke exotic stellar populations. Using cosmological radiation-hydrodynamic simulations with a standard chemical enrichment model, we identify a population of giant molecular clouds whose internal abundance patterns reproduce several key globular cluster signatures: large light-element abundance spreads and nitrogen–oxygen anticorrelations at nearly constant iron abundance. These clouds form at the restart of star-formation activity after an earlier starburst, where previously ejected oxygen-rich gas collides with nitrogen-rich galactic gas, and are sites of dense star-cluster formation. In this picture, the chemical abundance patterns of globular clusters need not require extended in-cluster star formation, but can be inherited at birth from chemically structured interstellar gas shaped by the baryon cycle. Globular clusters therefore provide a fossil record of chemical enrichment and gas flows in high-redshift galaxies.
    \end{mdframed}
\end{abstract}
\nokeywords

Globular clusters (GCs) represent some of the most ancient and dense stellar systems in the cosmos, with ages typically exceeding 10\,Gyr, present day masses of $\sim10^5-10^6\,\mathrm{M_\odot}$, and half-light radii of a few parsecs \citem{Harris:2013aa,Gratton:2019aa}. They are relics of the earliest star formation in the Universe, and are observed today across a wide range of galaxy stellar masses, making them excellent tracers of galactic structure formation \citem{Brodie:2006aa,Forbes:2018aa,Taylor:2025ab}. Yet despite being discovered over three centuries ago, their formation remains one of the foremost unsolved problems in astrophysics.

A defining feature of GCs is their chemical abundance patterns. Although most GCs show little dispersion in iron, they ubiquitously exhibit large star-to-star variations in light elements, such as He, C, N, O, and Na \citem{Carretta:2009aa,Gratton:2012aa,Bastian:2018aa,Gratton:2019aa}. These variations are characterised by anticorrelations, such as between [Na/Fe] and [O/Fe] and between [N/Fe] and [O/Fe]. These characteristics are unique to GCs and the prevailing explanation is that GCs host multiple populations of stars, formed over $\sim 1-100$\,Myr, that pollute the star-forming gas within the cluster with their ejecta. Proposed polluters include Asymptotic Giant Branch (AGB) stars \citem{DErcole:2010aa,Ventura:2013aa}, fast-rotating massive stars \citem{Decressin:2007aa}, interacting binary stars \citem{de-Mink:2009aa}, supermassive stars \citem{Denissenkov:2014aa,Gieles:2018aa} and extremely massive stars \citem{Gieles:2025aa}, but all such self-enrichment models face a severe mass-budget problem \citem{Bastian:2015aa,Bastian:2018aa} or rely on exotic and as-yet unobserved polluters \citem{Gieles:2025aa}.

Recent \textit{James Webb Space Telescope} (\textit{JWST}) observations have identified high-redshift galaxies with extreme nitrogen enhancement \citem{Bunker:2023aa,Isobe:2023aa}, reminiscent of the enriched stellar populations in GCs. This similarity has led to the suggestion that such systems may be sites of ongoing GC formation \citem{Senchyna:2024aa,Ji:2026aa,Naidu:2026aa}, with some models invoking the same exotic polluters proposed for GCs, including supermassive stars \citem{Charbonnel:2023aa}. However, \citetm{McClymont:2025ae} recently demonstrated that nitrogen-rich galaxies can arise naturally in high-redshift systems with a standard chemical enrichment model: stochastic star-formation histories promote the regular ejection and re-accretion of oxygen-rich gas, leaving AGB stars to periodically enrich the galactic gas with nitrogen. Here, we show that these same baryon-cycle processes also produce massive giant molecular clouds (GMCs) that match key chemical abundance patterns of GCs, including the N--O anticorrelation and spreads in Mg and Si. We highlight at the outset that these clouds were serendipitously discovered in a simulation designed to study galaxy formation rather than star clusters, and as such the simulation does not track all features relevant to GCs, including Na abundances and long-term star cluster evolution. Future work will be required to mature this scenario to the level of detail achieved for self-enrichment models over the last two decades.

We analyse the state-of-the-art THESAN-ZOOM suite of cosmological radiation-hydrodynamic simulations that follow galaxy evolution during the first two billion years of cosmic history \citem{Kannan:2025aa}. THESAN-ZOOM consists of a mixed-resolution set of simulations, reaching a maximum mass resolution of just 142\,$\mathrm{M_\odot}$ and a fiducial resolution of 1100\,$\mathrm{M_\odot}$. The simulation includes a ``standard'' chemical network that self-consistently tracks the enrichment of nine elements (H, He, C, N, O, Ne, Mg, Si, Fe), enabling us to resolve individual GMCs within their full cosmological environment (\methodslink). We identify GMCs throughout our simulation suite by searching for bound structures and select potential GC-like GMCs based on their chemical abundances. Specifically, we require large $\log\mathrm{(N/O)}$ spreads ($>0.5\,$dex), anticorrelated [N/Fe] and [O/Fe] (Pearson coefficient $<-0.3$), and a small dispersion in [Fe/H] ($\sigma_\mathrm{[Fe/H]}<0.1\,\mathrm{dex}$), consistent with the defining chemical properties of observed GCs (\methodslink). This approach isolates clouds that would imprint GC-like abundance patterns on the stars they form. 

\begin{figure*}
\centering
\includegraphics[width=\textwidth]{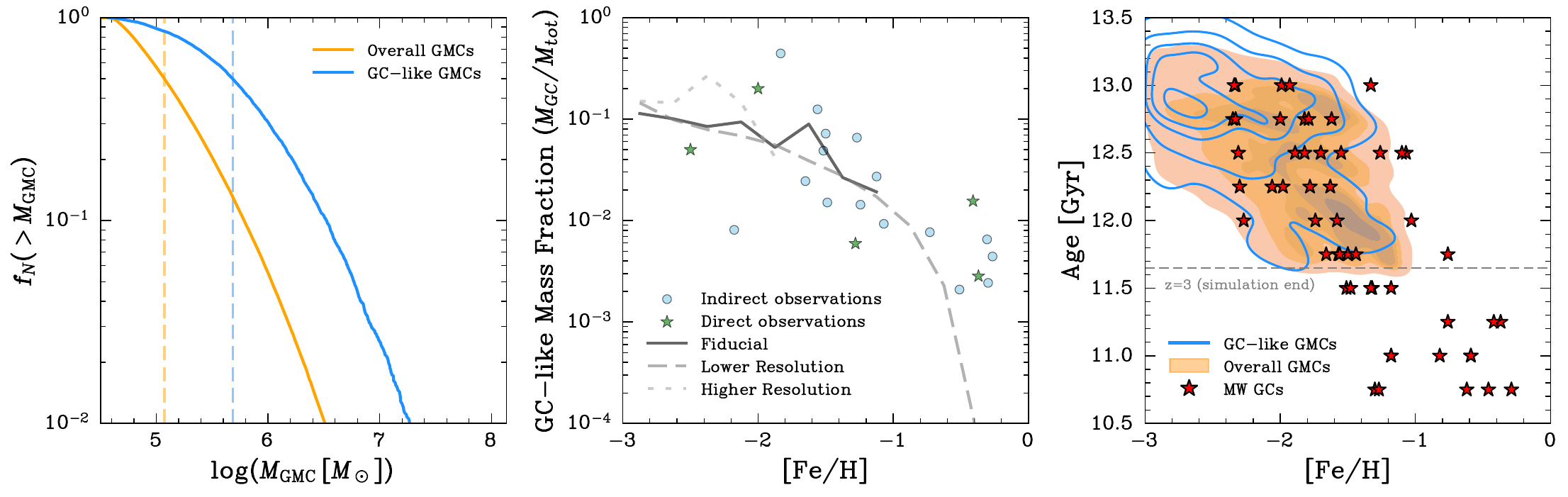}
\caption{\textbf{Population statistics of GC-like GMCs identified in our simulations.} Left: the normalized mass function of all GMCs compared to the GC-like GMCs, which are typically more massive, showing a median mass (dashed lines) that is $\sim0.5$\,dex higher. Centre: the mass of GC-like GMCs as a fraction of the total mass of all GMCs in bins of [Fe/H], averaged across all halos in the simulation. The mass fraction tends to decrease with metallicity, and is consistent with observations of the GC mass fraction observed in galaxies. Right: The age-metallicity distribution of simulated GC-like GMCs and overall GMCs as blue and orange contours, respectively (5th, 25th, 50th, 75th, and 95th percentiles). The AGB-cycling mechanism forms GC-like GMCs at sufficiently early times to match even the earliest forming MW GCs (red stars).}\label{fig1}
\end{figure*}

Figure~\ref{fig1} presents the population statistics of GC-like GMCs identified across the simulation suite. The left panel of Figure~\ref{fig1} compares the cumulative mass function of all GMCs to that of the GC-like GMCs. GC-like GMCs are typically more massive than ordinary GMCs, with a median mass $\sim0.5\,\mathrm{dex}$ higher. This is partially explained by the fact that more massive GMCs are able to sample a larger amount of the interstellar medium, and therefore are more likely to include gas with a variety of chemical abundances. As we demonstrate below, the higher masses also result from the specific formation conditions of the GC-like GMCs, which promote the formation of massive GMCs. The central panel shows the fraction of total GMC mass which is locked in GC-like GMCs as a function of [Fe/H]. The fraction decreases with increasing [Fe/H], falling from $\sim0.1$ at $\mathrm{[Fe/H]}=-3$ to $\sim0.01$ at $\mathrm{[Fe/H]}=-1$. The simulation is consistent with the observed fraction of stellar mass locked in GCs in nearby galaxies (\methodslink), indicating that the GC-like GMCs could form the observed number of GCs. The right panel demonstrates that GC-like GMCs are formed at sufficiently early times and low metallicities to be consistent with even the oldest MW GCs \citem{VandenBerg:2013aa}. We note that the simulated GMCs reproduce the relationship between $\log(\mathrm{N/O})$, [N/C], and [Mg/Si] spreads seen in MW GCs (Extended Data Figure~\ref{EDfig1}).

To identify the physical drivers of GC-like GMC formation, we employ a random forest classifier to distinguish GC-like GMCs from the general GMC population (\methodslink). GMC mass emerges as the strongest predictor, followed by formation redshift (Extended Data Figure~\ref{EDfig2}). Importantly, galactic-scale parameters, namely the star formation rate between 20 and 50\,Myr ago and the net gas inflow rate, are more predictive than local cloud metallicity. GC-like GMCs are preferentially more massive, form at higher redshift, and have lower metallicity, which are trends consistent with the observed properties of GCs \citem{Harris:2013aa,Gratton:2019aa} and confirm that our chemically selected GC-like GMCs also show the expected physical properties. Crucially, the importance of the galactic-scale parameters demonstrates that GC-like GMC formation depends on specific phases of the baryon cycle rather than cloud-scale properties alone. The variability of star formation and gas flows in high-redshift galaxies on these timescales is well-established both in observations \citem{Looser:2024aa,Witten:2025aa} and theory \citem{Tacchella:2020aa,McClymont:2025aa}.

\begin{figure}
\centering
\includegraphics[width=\columnwidth]{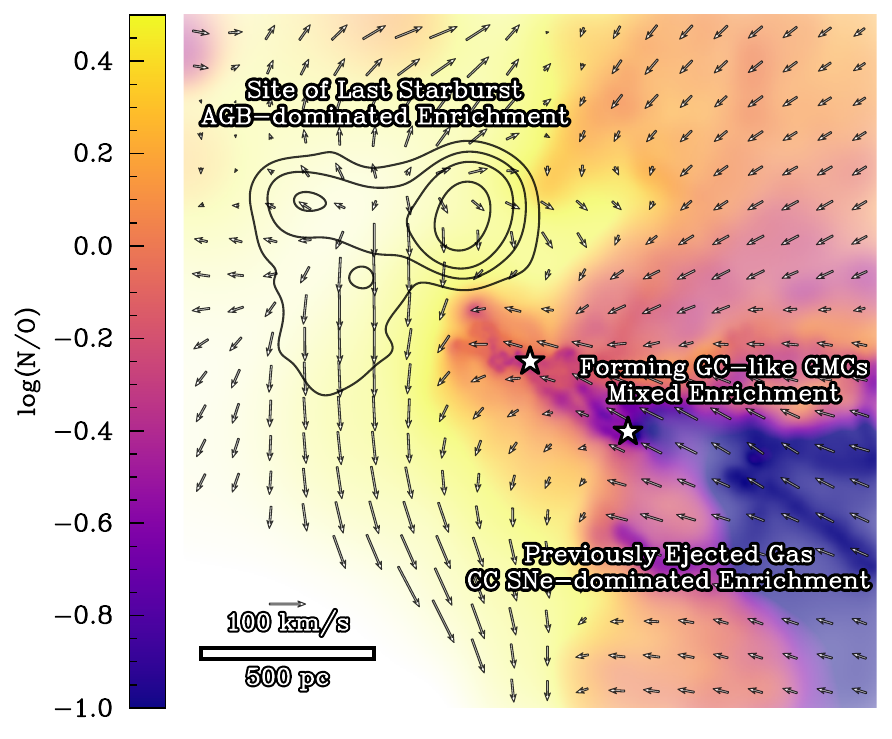}
\caption{\textbf{Inflow-driven formation of GC-like GMCs.} A 500\,pc depth slice through a galaxy at the onset of a new starburst. The colourmap shows mass-weighted log(N/O), with opacity set by the gas mass surface density. Overplotted are arrows showing the gas velocity perpendicular to the line of sight. White star markers showing the location of GC-like GMCs. The black contour shows the stellar mass distribution. Oxygen-rich gas (dark blue, shades of purple and red), ejected during the previous starburst, falls back onto the galaxy, colliding with and entraining the diffuse nitrogen-rich gas (yellow and shades of orange). This collision promotes the formation of massive GMCs with a large spread in their N and O abundances, which are the GC-like GMCs identified in this work.}\label{fig2}
\end{figure}

\begin{figure*}
\centering
\includegraphics[width=\textwidth]{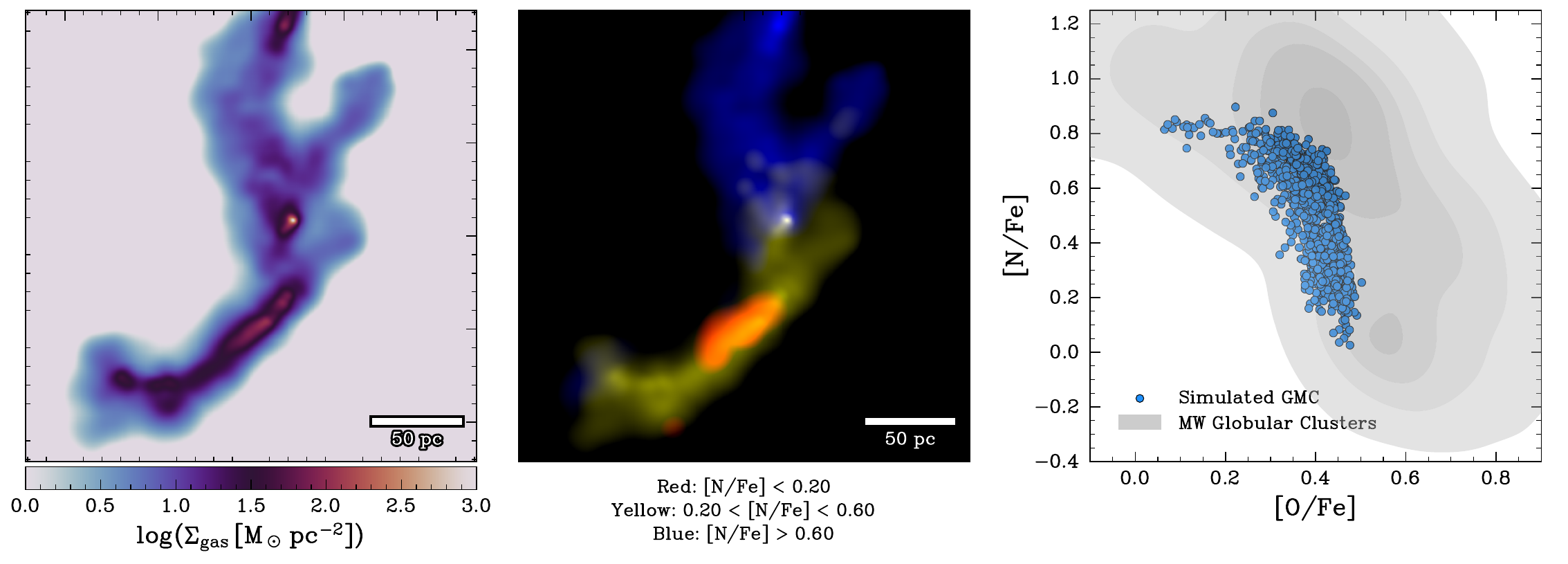}
\caption{\textbf{Resolved GMC with GC-like abundance structure.} Left: Gas surface density of a cloud formed during a re-accretion event onto a temporarily quenched galaxy at $z=7.3$. Centre: RGB composite highlighting spatial variations in nitrogen abundance, showing two chemically distinct diffuse streams converging into a dense central core. Right: [N/Fe] versus [O/Fe] for individual gas cells in the cloud (blue), compared to stellar abundances of the Milky Way globular clusters shown as grey contours (5th, 25th, 50th, 75th, and 95th percentiles). The cloud exhibits the pronounced N-O anticorrelation and minimal iron dispersion ($\sigma_\mathrm{[Fe/H]}=0.05\,\mathrm{dex}$) characteristic of globular clusters, indicating that the abundance pattern is established prior to cluster formation.
}\label{fig3}
\end{figure*}

\begin{figure}
\centering
\includegraphics[width=\columnwidth]{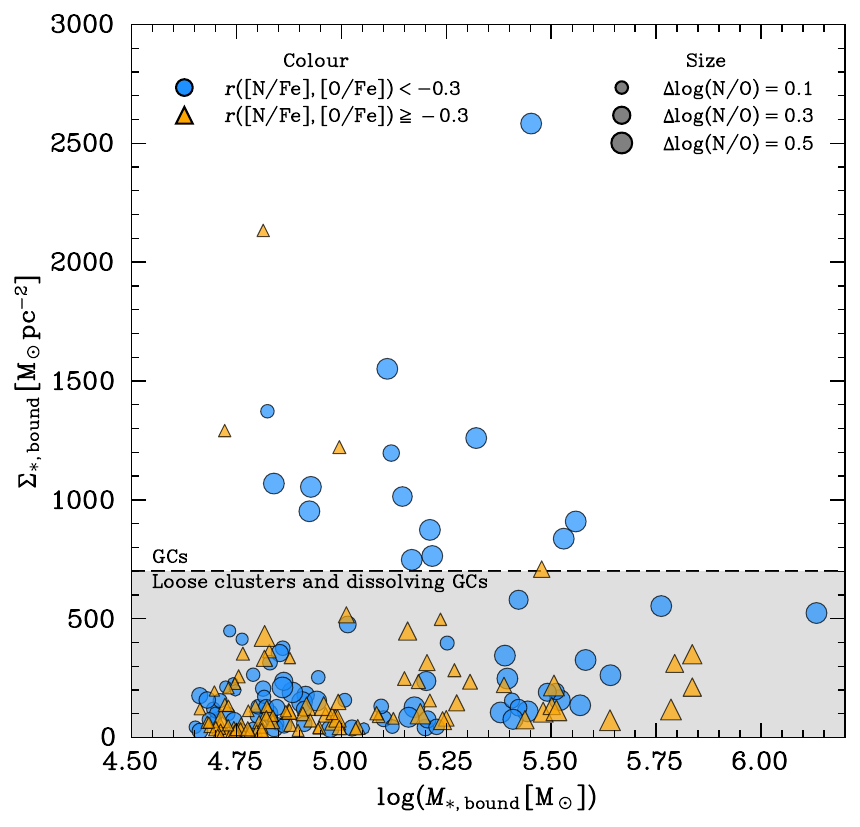}
\caption{\textbf{Connecting chemical abundances to dense star cluster formation.} Bound stellar surface density of clusters versus their bound stellar mass. Orange triangles (blue circles) denote systems with [N/Fe]–[O/Fe] (anti)correlations. The marker size is proportional to the $\log(\mathrm{N/O})$ spread in the birth gas. Massive clusters ($M_{\ast,\mathrm{bound}}\gtrsim10^5\,\mathrm{M_\odot}$) above $\Sigma_\ast \approx 700\,\mathrm{M_\odot\,pc^{-2}}$, comparable to observed globular cluster surface densities, exhibit anticorrelations and significant abundance spreads. Lower-density systems show more diverse abundance behaviour, and consist of a mixture of dissolving GCs and intrinsically loosely-bound star clusters. Dense, low-mass clusters tend to exhibit minimal spreads, providing tentative evidence for cluster mass-dependent spreads. These results support a link between dense star cluster formation and GC-like chemical abundances in the parent GMC.
}\label{fig4}
\end{figure}

The formation channel is driven by the interaction between galactic inflows and a chemically structured interstellar medium (Fig.~\ref{fig2}). A starburst first expels oxygen-rich core-collapse supernova ejecta out of the shallow potential well of the dwarf galaxy. As the galaxy temporarily quenches, intermediate-mass AGB stars from older populations continue to enrich the remaining gas with nitrogen. The ejected oxygen-rich gas subsequently cools in the halo and falls back onto the galaxy. The collision of this high-velocity, oxygen-rich inflow with the diffuse, nitrogen-rich gas creates a zone of extreme compression. This collision promotes the rapid gravitational collapse of massive GMCs that contain a mixture of the abundances found in the diffuse gas and the inflowing gas, naturally producing GMCs with the large $\log\mathrm{(N/O)}$ spreads and [N/Fe]--[O/Fe] anticorrelations characteristic of observed GCs. While this inflow-driven channel dominates, we also find some GC-like GMCs forming through other compression channels, such as positive feedback. 

Figure~\ref{fig3} shows a GC-like GMC identified in our highest-resolution simulation, illustrating how the galactic-scale process is imprinted on the scale of an individual cloud. This GC-like GMC formed in an inflow to a mini-quenched galaxy at redshift $z=7.3$, at the onset of a new starburst. The cloud structure reveals two diffuse arms with distinct chemical abundances that feed into a dense central core. The core reaches high gas surface densities ($\Sigma_{\rm gas} >10^3\,\mathrm{M_\odot\,pc^{-2}}$), comparable to the dense, compressed environments in which previous high-resolution simulations of gas-rich dwarf galaxies form massive bound clusters \citem{Lahen:2019aa,Lahen:2020aa}, and, crucially, contains a mixture of nitrogen-rich and nitrogen-poor gas. The distribution of [N/Fe] versus [O/Fe] shows the anticorrelation characteristic of observed GCs. For comparison, we plot with contours the stellar abundances of MW GCs (\methodslink), which show similar abundance patterns. A star cluster forming from this gas would therefore inherit this chemical spread at birth, without the need for multiple distinct epochs of star formation separated by millions of years.

To test whether chemically selected GC-like GMCs produce bona fide dense clusters, we identify bound star clusters in a subset of simulations optimized for cluster-scale structure formation (\methodslink). Figure~\ref{fig4} shows the physical and chemical properties of the identified star clusters. The plot shows the bound stellar mass surface density of star clusters plotted against their bound stellar mass, with the size of the points representing the spread of log(N/O) in their birth gas. Massive ($M_{\ast,\mathrm{bound}}>10^5\,\mathrm{M_\odot}$) star clusters with $\Sigma_\ast\gtrsim700\,\mathrm{M_\odot\,pc^{-2}}$ show anticorrelated [N/Fe]-[O/Fe] ($r\leq-0.3$) and at least a moderate spread in log(N/O) ($\Delta\log(\mathrm{N/O})_\mathrm{bound}\geq 0.2$). Here we have defined the surface density threshold to split the chemical behaviour of the star clusters, yet it is broadly consistent with the surface densities observed in GCs \citem{Baumgardt:2018aa} and with the threshold for forming bound massive clusters predicted by numerical simulations \citem{Fukushima:2023aa}. We stress that the exact value of this threshold may be impacted by our resolution, yet the existence of this threshold supports a direct link between dense star cluster formation and GC-like abundances. Star clusters below the density threshold show a variety of spreads and (anti)correlations, consistent with a mixture of dissolving GCs and intrinsically looser star clusters that lack GC-like abundance patterns. There are four low-mass clusters above the density threshold which show a very small log(N/O) spread, indicating that some low-mass GCs may not contain observable abundance spreads, although these clusters may be below the minimum initial mass of surviving MW GCs accounting for dynamical mass loss. This also provides tentative evidence that more massive GCs formed in our proposed scenario would have larger spreads and greater chemical complexity. Such a trend is seen in observations \citem{Milone:2017aa,Dondoglio:2025aa}, as well as in our simulated GMCs (Extended Data Figure~\ref{EDfig3}).

Several aspects require further investigation. Our simulations do not track all elements relevant to GC abundance patterns. In particular, we do not track sodium, and so we cannot directly verify whether the [Na/Fe]--[O/Fe] anticorrelation is reproduced. However, sodium is produced alongside nitrogen in AGB stars through the NeNa cycle \citem{Karakas:2010aa}, and so we expect that the same processes which create the [N/Fe]--[O/Fe] anticorrelation would also produce the [Na/Fe]--[O/Fe] anticorrelation. Indeed, there is a rich body of work which demonstrates that AGB stars in the self-polluter scenario can reproduce the chemical features of globular clusters, including the [Na/Fe]--[O/Fe] anticorrelation \citem{DErcole:2008aa,Ventura:2008aa,Ventura:2009aa,DErcole:2010aa,DAntona:2016aa}. Additionally, future simulations that are capable of directly modelling the evolution of dense star clusters across cosmological timescales and in large volumes will be essential to confirm that this model produces the correct abundance and properties of GCs at $z=0$.

Our results have a number of important implications for the interpretation of GC abundance patterns and observations of high-redshift galaxies. Under this scenario, when we observe highly nitrogen-rich high-redshift galaxies with the \textit{JWST}, we are indeed likely witnessing the formation of globular clusters because the processes that generate nitrogen-rich galaxies and nitrogen-rich star clusters are the same. This may not be the case for mildly nitrogen-rich galaxies, which may be the result of an insufficiently potent burst-lull cycle to promote GC formation (or the originally higher nitrogen abundances may have already been diluted by self-enrichment). 

Our scenario predicts a very short total formation timescale for GCs of $\lesssim3-5$\,Myr, constrained both by the short dynamical times of the GMCs and the requirement that star formation should cease before core-collapse supernovae inject iron, which would introduce a spread in [Fe/H]. Star formation could be more extended in iron-complex GCs. This rapid formation timescale is consistent with the inferred small age spreads of globular clusters \citem{Bastian:2013ab,Martocchia:2018aa}. 

An important advantage of this scenario is that it avoids the mass-budget problem that tends to plague self-enrichment scenarios. In our model, so-called 1G and 2G stars are formed simultaneously and from pre-enriched gas. As the gas was already pre-enriched, there is no need for the 1G population to be many times larger than the 2G population. As such we do not need to invoke huge mass losses of GCs, although we do not need to preclude it either. Furthermore, we do not expect preferential loss of either nitrogen-rich or nitrogen-normal stars, unlike self-enrichment scenarios that typically require the 1G population to be preferentially stripped. This is because under our scenario there is no obvious reason why nitrogen-rich or nitrogen-normal stars should have different radial distributions, although there may be cluster-to-cluster variation. This is consistent with observations, where some GCs show radial gradients in their population ratios, but others do not \citem{Dalessandro:2019aa,Leitinger:2023aa}.

In our scenario, it is possible that nitrogen-rich stars may form in the field. The diffuse nitrogen-rich gas spans across a mini-quenched galaxy, and so some nitrogen-rich stars could form at the very onset of a starburst even outside of GC-like GMCs. This is significant because nitrogen-rich field stars are commonly used as tracers of dissolved GCs \citem{Martell:2011aa,Schiavon:2017aa,Belokurov:2023aa,Kane:2025aa}. Determining the fraction of nitrogen-rich stars formed (and remaining) within GCs predicted by our scenario will require simulations that trace stellar abundances and that can trace the evolution of GCs over cosmological timescales. 

While further theoretical development of the model is clearly required for detailed quantitative predictions, there are several avenues to test this scenario in the near future. For example, our predictions that even very young ($\lesssim$1\,Myr) high-redshift forming GCs should show anomalous chemistry and that forming GCs should not have star formation extended beyond $\sim3-5$\,Myr can be directly tested by measuring the ages and chemical abundances of GC-progenitors. This has been possible for a small number of gravitational lensed systems \citem{Pascale:2023aa,Pascale:2024aa}, but a larger sample including even younger clusters is required to be constraining. Additionally, our scenario predicts that (low-mass, high-redshift) galaxies which have been quenched for $\gtrsim50$\,Myr should show GC-like chemical abundances, including nitrogen enhancement. This prediction is in principle testable with measurements of absorption lines or residual line emission, yet due to the faint nature of these sources, it may require lensed sources or stacking techniques.

More broadly, if GC abundance patterns are indeed inherited from GMCs rather than the result of internal processes, then GCs represent an unparalleled probe of the chemical and dynamical state of the high-redshift interstellar medium. Detailed chemical abundance studies of GCs could therefore provide powerful new constraints on the baryon cycle and chemical enrichment of the first galaxies, complementing the direct observations currently underway with the \textit{JWST}.

\clearpage
\bibliographym{mainbib}

\clearpage
\makeatletter
\renewcommand{\fnum@figure}{Extended Data Figure \thefigure}
\renewcommand{\fnum@table}{Extended Data Table \thetable}
\makeatother
\setcounter{figure}{0}
\setcounter{table}{0}
\setcounter{footnote}{0}

\newpage

\phantomsection
\label{sec:methods}
\section*{Methods}

\subsection*{Simulations}

We use the THESAN-ZOOM suite of cosmological radiation-hydrodynamics simulations \citem{Kannan:2025aa}. These are zoom-in resimulations of halos selected from the larger THESAN volume \citemet{Kannan:2022aa, Smith:2022ab, Garaldi:2022aa}, designed to study the formation of galaxies during the Epoch of Reionization. The simulations are run using the moving-mesh code AREPO \citemet{Springel:2010aa,Kannan:2019aa,Zier:2024aa}, which employs a finite-volume Godunov scheme on an unstructured Voronoi mesh. This approach allows for high dynamic range and quasi-Lagrangian adaptivity, essential for resolving the dense interstellar medium within high-redshift galaxies. The suite is composed of three distinct resolution levels, referred to in this paper as ``Lower Resolution'', ``Fiducial'', and ``Higher Resolution'', corresponding to spatial resolutions 4, 8 and 16 times finer than in THESAN and to baryonic particle masses of $9.09\times10^3\,\mathrm{M}_\odot$, $1.14\times10^3\,\mathrm{M}_\odot$, and $1.42\times10^2\,\mathrm{M}_\odot$ respectively. The minimum gravitational softening lengths for gas are 69.2\,cpc, 34.6\,cpc, and 17.3\,cpc for each resolution level respectively, whereas the stars and dark matter have gravitational softening lengths of 553.59\,cpc, 276.79\,cpc, and 138.39\,cpc. Thus while we can effectively resolve GMCs in our simulations, the long-term dynamical evolution of dense star clusters is below the resolution limit.

The physics model includes non-equilibrium thermochemistry, coupled dust formation and destruction, and a comprehensive stellar feedback model accounting for winds, core-collapse supernovae, and Type Ia supernovae. Crucially, the simulations track the production and advection of nine distinct chemical elements (H, He, C, N, O, Ne, Mg, Si, Fe), using mass and metallicity-dependent yield tables for AGB stars \citemet{Karakas:2010aa,Doherty:2014aa,Fishlock:2014aa}, core-collapse supernovae \citemet{Portinari:1998aa,Kobayashi:2006aa}, and Type-Ia supernovae \citemet{Nomoto:1997aa}. This enrichment scheme was implemented for the IllustrisTNG simulations \citemet{Pillepich:2018aa} and adapted from the Illustris enrichment scheme \citemet{Vogelsberger:2013aa}. 

In our analysis we include the m13.0, m12.6, m12.2, and m11.5 lower resolution simulations, the m11.1, m10.8, m10.4, m10.0 fiducial resolution simulations, and the m9.7, m9.3, m8.9, m8.5, and m8.2 high-resolution simulations (see Table~3 in \citetm{Kannan:2025aa} for details). We use additional fiducial-resolution simulations of m10.4 and m9.7, which were run without the additional empirically calibrated early stellar feedback channel. We use these simulations exclusively to search for bound star clusters, the formation of which is more accurately modelled in these simulations as they only include physically motivated feedback that is less effective in disrupting GMCs. This change does not directly impact the chemical abundances injected into the gas.

The number in the name of each simulation refers to the logarithmic mass of the parent simulation object at $z=3$, which is the final redshift of the simulations. Halos were identified with the friends-of-friends (FoF) algorithm \citemet{Davis:1985aa}.

\subsection*{GMC identification and GC-like GMC selection}

We identify giant molecular clouds (GMCs) using an implementation of the CloudPhinder code \citemet{Guszejnov:2020aa}. CloudPhinder identifies bound, self-gravitating structures. It operates by first constructing a density hierarchy of the gas and then iteratively descending this hierarchy to locate substructures that are gravitationally bound (virial parameter $\alpha_\mathrm{vir}<\alpha_\mathrm{max}$). We select $\alpha_\mathrm{max}=20$ in order to capture loosely bound clouds, though our analysis is not sensitive to this choice. The median virial parameter of GMCs considered in this work in our fiducial resolution simulations is 4.26.

We include cold ($T<1000$\,K), relatively dense ($n_\mathrm{H}>1\,\mathrm{cm}^{-3}$) gas and young stars ($<3$\,Myr) in the algorithm. We impose a strict stellar mass fraction cut of $M_\ast/M_\mathrm{gas}<0.01$ to ensure the GMC abundances have not been strongly affected by ongoing star formation. We only include well-resolved clouds ($N_\mathrm{gas}>30$\,cells). For each identified cloud, we compute mass-weighted chemical abundances and kinematic properties. To ensure our analysis is relevant to observed MW GCs and to avoid sensitivity to primordial enrichment, we further impose an average metallicity cut of $\langle$[Fe/H]$\rangle_{\mathrm{GMC}}>-3$.

\begin{figure*}
\centering
\includegraphics[width=\textwidth]{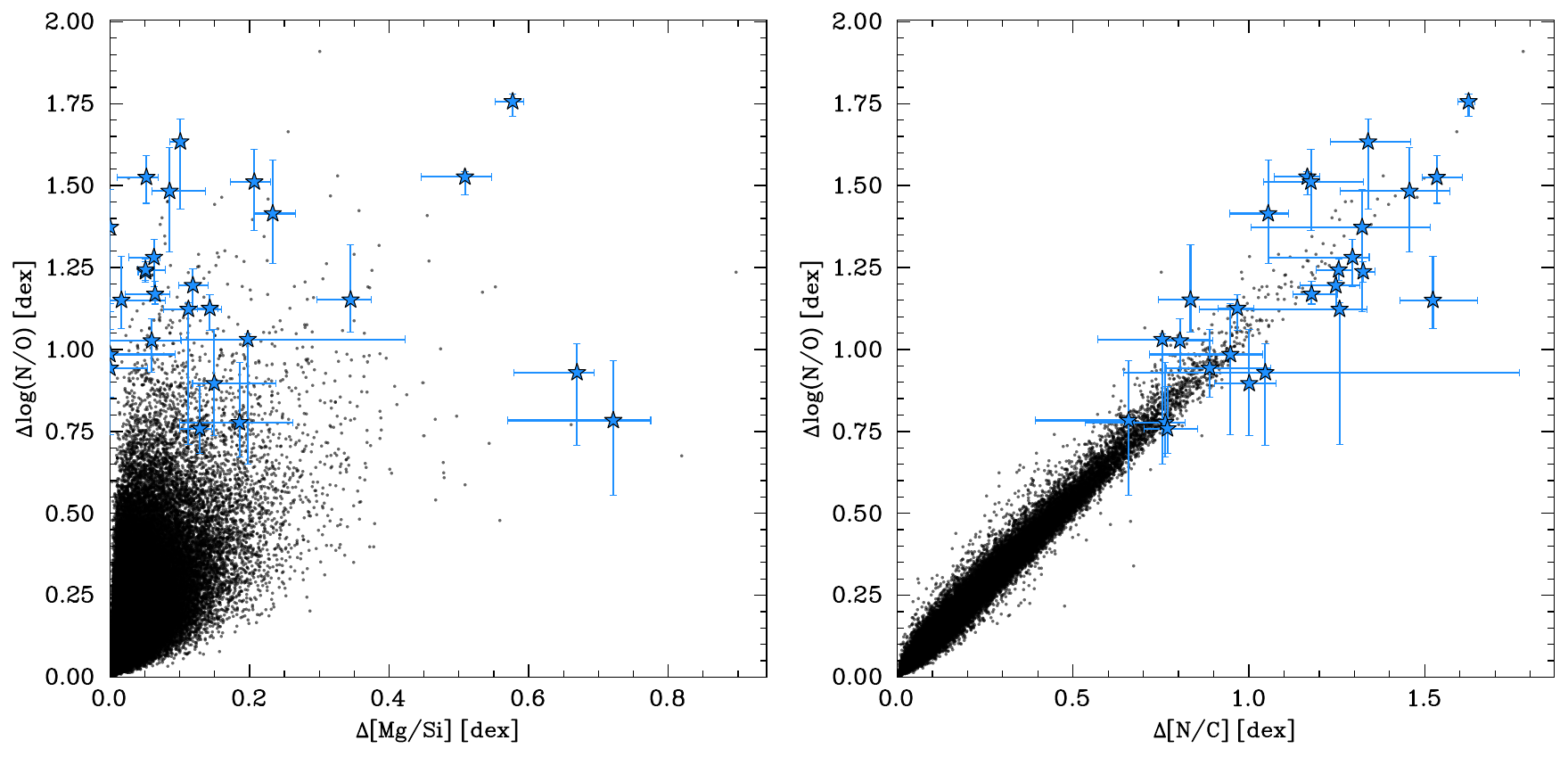}
\caption{\textbf{Chemical spread distribution in GMCs.} Simulated GMCs selected in our fiducial resolution simulations with anticorrelated [N/Fe] and [O/Fe] and a small dispersion in [Fe/H] are shown as black points. We include GMCs from all resolution levels to maximize visibility of the distribution in the tails. MW GCs are shown as blue stars. Left: GC-like GMCs (defined as $\Delta\log(\mathrm{N/O})>0.5$) show a wide variety of [Mg/Si] spreads, with no clear trend with $\log(\mathrm{N/O})$ spread. The same is true of MW GCs. Right: The spread in $\log(\mathrm{N/O})$ compared to the spread in [N/C]. In both the simulated GMCs and MW GCs, the distribution is dominated by the slope with relatively small scatter, indicating that carbon and oxygen abundances are tightly linked. This is important to reproduce for our AGB scenario, which one may naively expect could produce excessive carbon spreads.}\label{EDfig1}
\end{figure*}

\begin{figure*}
\centering
\includegraphics[width=\textwidth]{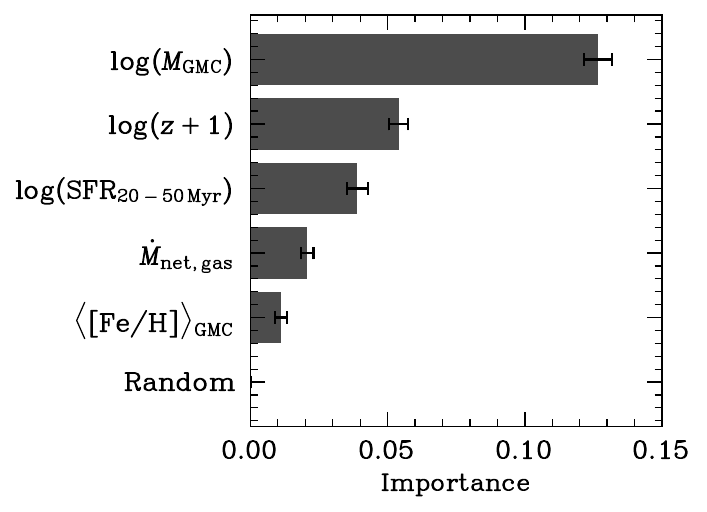}
\caption{\textbf{Importance of features to identify GC-like GMCs.} Relative feature importance from a random forest classifier distinguishing GC-like GMCs from the general GMC population. GMC mass and redshift are the strongest predictors, followed by the recent (previous 20–50 Myr) star formation rate and net gas inflow rate, indicating that GC formation is regulated by galaxy-scale baryon-cycle phases rather than cloud-scale properties alone. Error bars show the standard deviation across cross-validation folds; a random feature is included for reference.}\label{EDfig2}
\end{figure*}

\begin{figure*}
\centering
\includegraphics[width=0.9\textwidth]{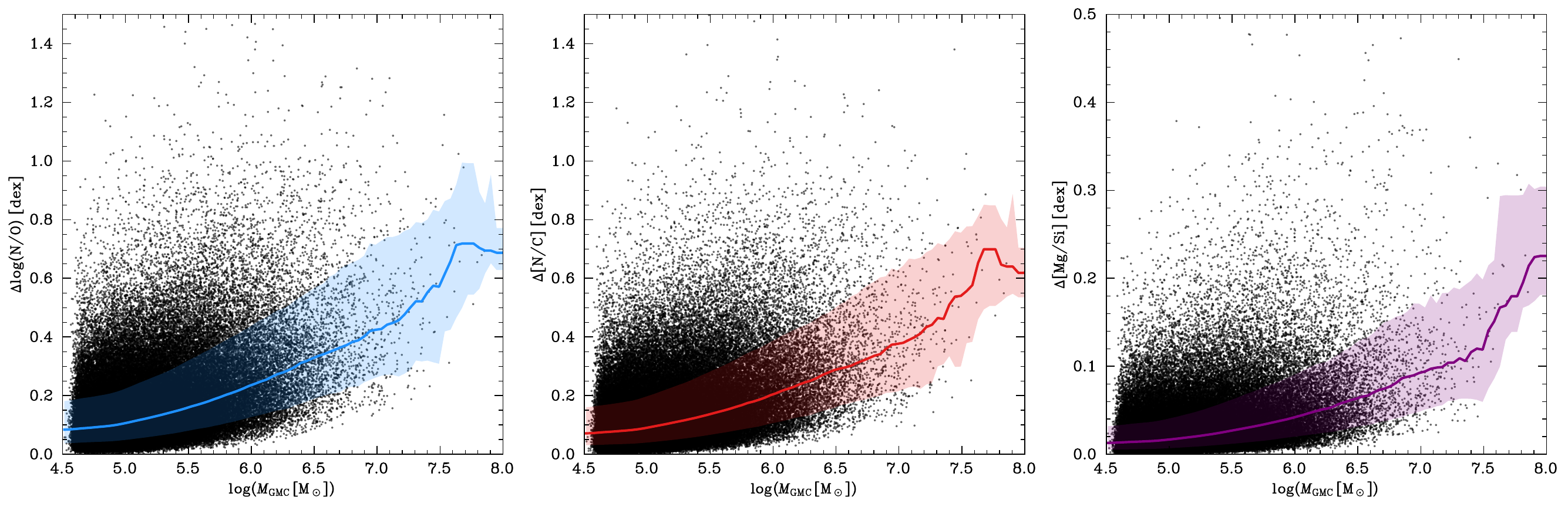}
\caption{\textbf{Chemical spreads dependent on GMC mass.} The $\log(\mathrm{N/O})$, [N/C], and [Mg/Si] spreads as a function of mass for GMCs selected in our fiducial resolution simulations with anticorrelated [N/Fe] and [O/Fe] and a small dispersion in [Fe/H]. Spreads tend to increase with increasing GMC mass. If this translates to a trend with the star cluster mass (as is tentatively indicated in Figure~\ref{fig4}), then this scenario would reproduce the trend seen in observed GCs.}\label{EDfig3}
\end{figure*}

\begin{figure*}
\centering
\includegraphics[width=\textwidth]{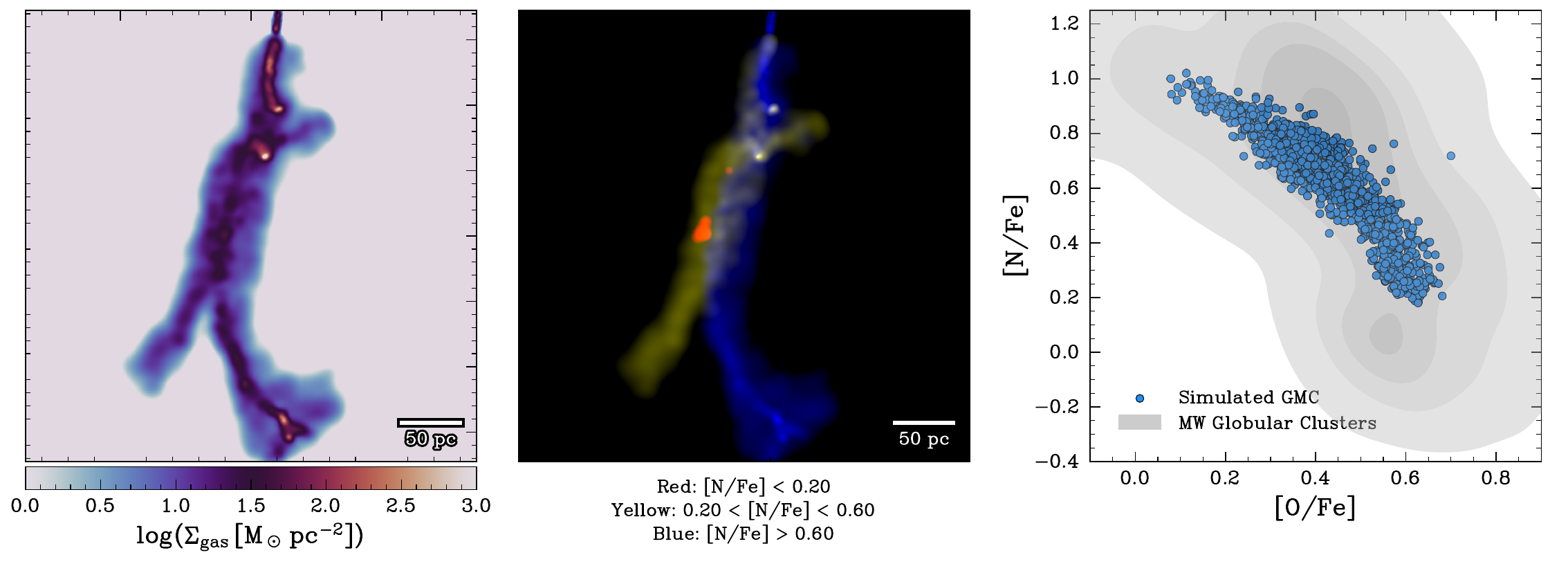}
\caption{\textbf{A GMC identified in our high-resolution simulation but not selected as a GC-like GMC.} The same as Figure~\ref{fig3} for a different GMC. This GMC is not selected as a GC-like GMC due to a large iron spread ($\sigma_\mathrm{[Fe/H]}=0.15\,\mathrm{dex}$).}\label{EDfig4}
\end{figure*}

\begin{figure*}
\centering
\includegraphics[width=\textwidth]{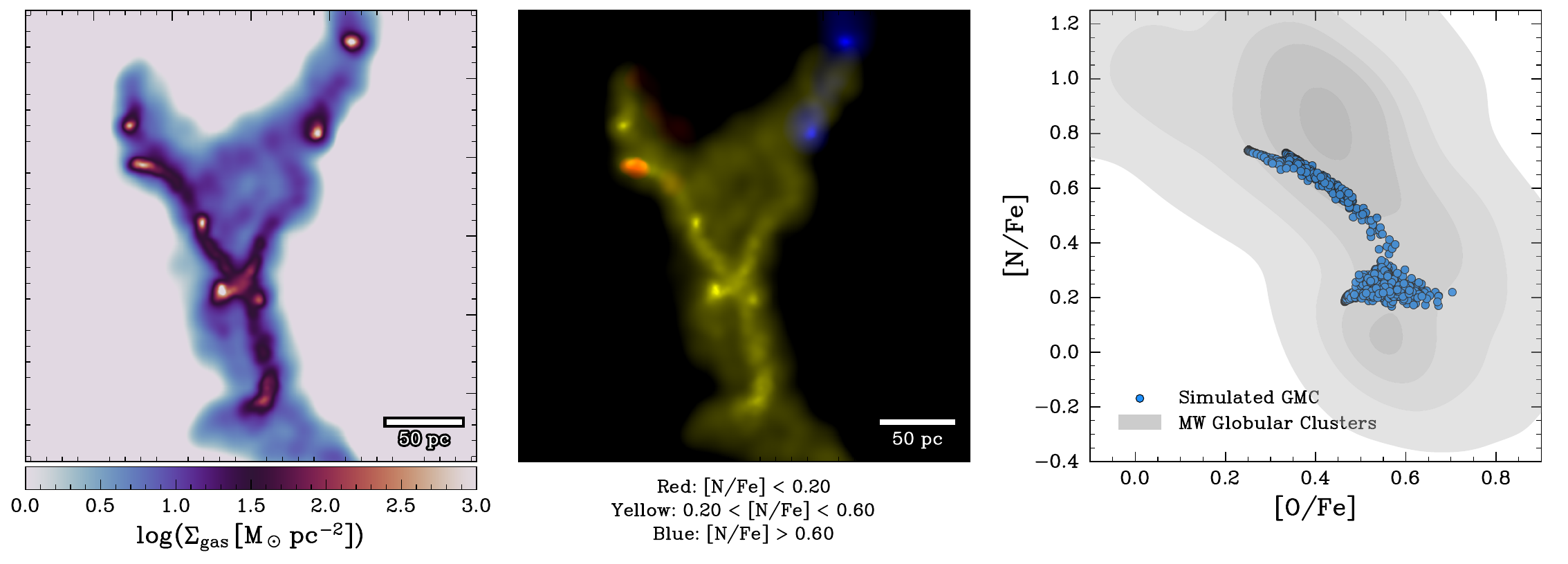}
\caption{\textbf{Additional example of a GC-like GMC identified in our high-resolution simulation.} The same as Figure~\ref{fig3} for a different GMC.}\label{EDfig5}
\end{figure*}

\begin{figure*}
\centering
\includegraphics[width=\textwidth]{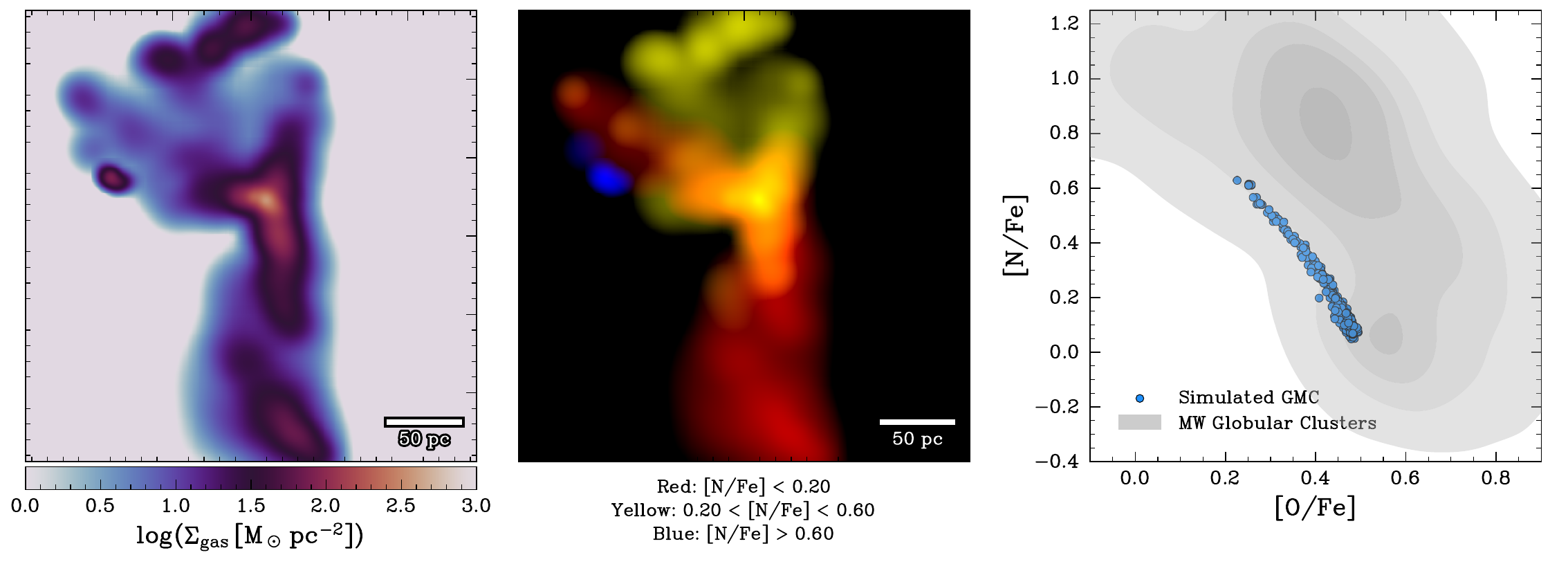}
\caption{\textbf{A GC-like GMC identified in our fiducial-resolution simulation.} The same as Figure~\ref{fig3} for a GMC in our fiducial-resolution simulation.}\label{EDfig6}
\end{figure*}

\begin{figure*}
\centering
\includegraphics[width=\textwidth]{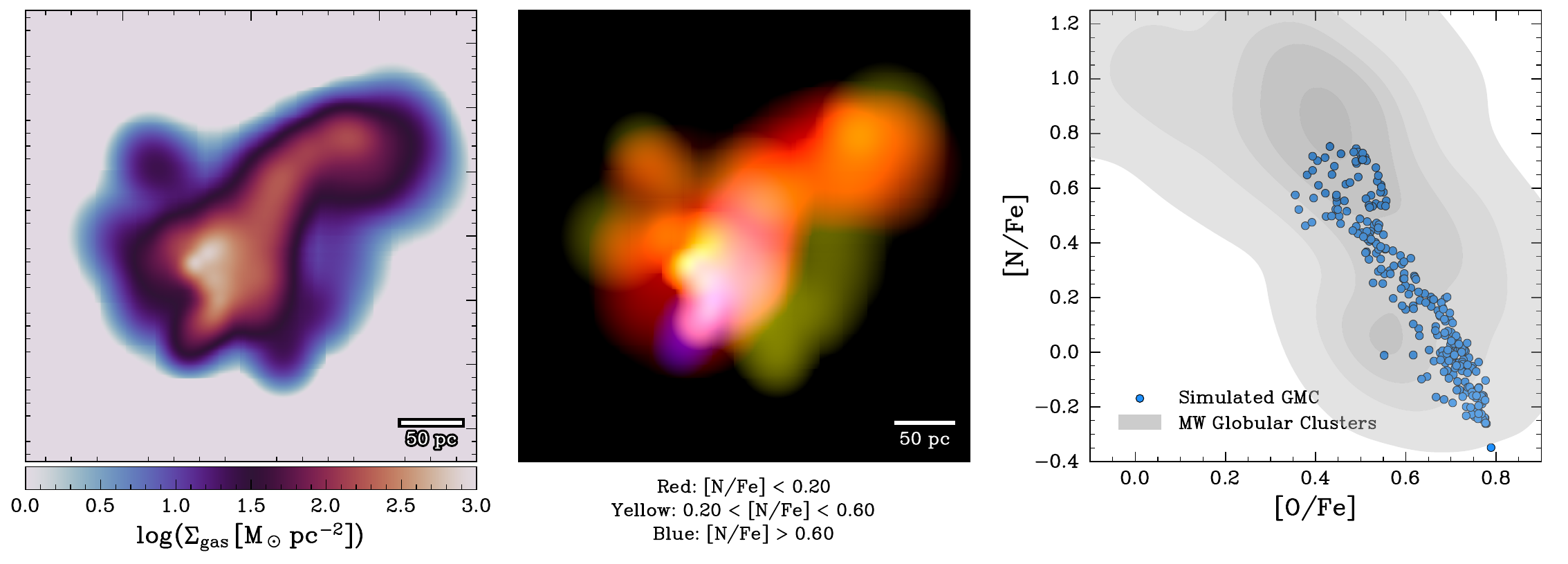}
\caption{\textbf{A GC-like GMC identified in our lowest-resolution simulation.} The same as Figure~\ref{fig3} for a GMC in our lowest-resolution simulation.}\label{EDfig7}
\end{figure*}

To identify potential GC-like GMCs, we apply chemical selection criteria derived from observational definitions of local GCs. The range of the $\log\mathrm{(N/O)}$ distribution within the cloud must exceed 0.5\,dex, indicating a significant spread in light element abundances. The standard deviation of [Fe/H] must be less than 0.1\,dex, ensuring the object is chemically homogeneous in heavy elements. The Pearson correlation coefficient between [N/Fe] and [O/Fe] must be negative ($<-0.3$), reflecting the characteristic anticorrelation. Our definition is chosen to be consistent with the chemical properties of the Type I GC population \citem{Dondoglio:2025aa}. The specific $\log\mathrm{(N/O)}$ threshold of 0.5\,dex was selected to be approximately consistent with the lower bound of uncertainty on the smallest spread we measured for MW GCs (Extended Data Figure~\ref{EDfig1}), which is $\sim$0.6. Selecting based on [N/Fe] spread instead of $\log\mathrm{(N/O)}$ yields similar results. This definition excludes the progenitors of iron-complex GCs. One such example is shown in Extended Data Figure~\ref{EDfig4}. Further examples of GC-like GMCs identified in different resolution levels are shown in Extended Data Figures~\ref{EDfig5},~\ref{EDfig6},~and~\ref{EDfig7}.

\subsection*{Star cluster identification}

We identify bound star clusters in the simulation using an implementation of the Phinder code \citemet{Grudic:2018aa}. Phinder identifies local minima in the gravitational potential field of stars and then identifies bound star clusters. As the stellar gravitational softening length is too large to accurately model the long-term evolution of dense star clusters, we instead search for young star clusters that would have been bound as they formed. We therefore use the initial stellar particle masses and the minimum gravitational softening length of gas, and we require that the median age of stars in a cluster is less than 10\,Myr to ensure this technique is appropriate. To avoid double-counting clusters across snapshots, we check whether any clusters share $\gtrsim$25\% of stellar particles with the same ID and retain only the youngest star cluster.

As the simulation does not track the chemical abundances of stars, we instead consider the chemical abundances of the gas from which the stars form. To do this, we find the snapshot prior to the formation of the oldest star in the cluster (usually the previous snapshot), and match the IDs of gas in that snapshot to those of the stars which later form. We found that in some cases the birth gas was already being contaminated by ongoing cluster formation due to cluster stars which were not identified as bound at the snapshot where the cluster was identified. To address these cases, we search for young stars near the birth gas which could have already significantly polluted the birth cloud. Specifically, we search for stars between 2--5\,Myr old which are closer to a birth gas cell than the 98th percentile inter-gas-cell spacing (capped at 50\,pc), and if any such stars are identified then instead search the previous snapshot. This process accurately captures the chemical abundances of the gas that will go on to form the bound star cluster. We require at least 10 matched birth gas cells in order to include a cluster in our analysis and only consider clusters with an average birth gas metallicity of $\langle$[Fe/H]$\rangle_{\mathrm{SC}}>-3$. In Figure~\ref{fig4}, we exclusively consider the properties of the identified bound stellar particles and their associated birth gas in order to clearly delineate bona-fide dense star clusters.

\subsection*{Machine learning analysis}

To isolate the physical drivers of the formation of GC-like GMCs, we employ a Random Forest approach using the scikit-learn Python package. We train a Random Forest Classifier to distinguish GC-like GMCs from the general GMC population on the fiducial resolution simulations. We evaluated models using the Receiver Operating Characteristic Area Under the Curve (ROC AUC) score. To ensure robust feature selection, we utilize Out-of-Fold (OOF) Permutation Importance. The model evaluation is performed using Repeated Stratified K-Fold cross-validation (10 splits repeated 10 times) to mitigate the effects of dataset variance and class imbalance. The ROC AUC score of the model is 0.913.

To select hyperparameters, we performed a randomised search with 60 parameter combinations using 3-fold cross validation. This led us to use 50 estimators, a maximum depth of 6, 13 minimum samples to split, and 6 minimum samples per leaf. 

We limit the GMCs considered in the analysis to within $0.2R_\mathrm{vir}$ in order to fairly assess the impact of galactic properties by focusing on the central galaxy within a halo. This gives a sample of 137,124 total GMCs, of which 2,488 are GC-like GMCs. Stellar masses, SFRs, and net gas inflow rates are calculated within $0.2R_\mathrm{vir}$ of the halo. We considered SFRs across bins of 0, 10, 20, 50, 100, 250, and 500\,Myr. We repeated the Random Forest analysis, eliminating SFR bins (and stellar mass) until the most important bin remained; 20--50\,Myr. 

\subsection*{Observational Data}

We plot the present-day globular-cluster stellar-mass fraction,
\begin{equation}
f_{\rm GC} \equiv \frac{M_{\rm GCS}}{M_\star} \, ,
\end{equation}
against host metallicity, $[{\rm Fe/H}]$, for dwarf galaxies with measured globular-cluster-system (GCS) masses. The host stellar masses, $M_\star$, and GCS masses, $M_{\rm GCS}$, are taken from the homogenized dwarf-galaxy GCS compilation of \citetmet{Dornan:2026aa}, which combines measurements from ACS Virgo \citemet{Peng:2008aa}, ACS Fornax \citemet{Liu:2019aa}, the Fornax Deep Survey \citemet{Prole:2019aa}, ELVES \citemet{Carlsten:2022aa}, MATLAS \citemet{Marleau:2024aa}, and nearby-dwarf samples including \citetmet{Georgiev:2010aa} and \citetmet{Gannon:2024aa}.

The metallicity sample is split into two groups:
\begin{enumerate}
\item \textit{Local dwarfs} (5 objects): Sagittarius dSph, LMC, WLM, Fornax dSph, and Eridanus II. We adopt host metallicities from Sagittarius dSph \citemet{Monaco:2005aa}, LMC \citemet{Choudhury:2016aa}, WLM \citemet{Leaman:2013aa}, Fornax dSph \citemet{de-Boer:2016aa} (with GC-system context from \citetmet{Larsen:2012aa}), and Eridanus II \citemet{Fu:2022aa} (with GC-system context from \citetmet{Li:2017aa,Orkney:2022aa}).

\item \textit{Other dwarfs} (12 objects): five Fornax-cluster dwarfs from \citetmet{Romero-Gomez:2023aa}, five ultra-diffuse galaxies (UDGs) from \citetmet{Gannon:2024aa}, and two nearby dwarfs (IKN and NGC~4163) from the Local Volume Database \citemet{Pace:2025aa}.
\end{enumerate}

For the \citetmet{Romero-Gomez:2023aa} and \citetmet{Gannon:2024aa} subsamples, $[{\rm Fe/H}]$ is derived from published $[{\rm M/H}]$ and $[\alpha/{\rm Fe}]$ using the \citetmet{Salaris:1993aa} relation,
\begin{equation}
[{\rm Fe/H}] = [{\rm M/H}] - \log_{10}\left(0.638 \times 10^{[\alpha/{\rm Fe}]} + 0.362\right).
\end{equation}
Uncertainties in $[{\rm Fe/H}]$ are propagated from the quoted $[{\rm M/H}]$ and $[\alpha/{\rm Fe}]$ errors.

We apply a uniform quality cut requiring finite metallicity uncertainty and $\sigma_{[{\rm Fe/H}]} < 0.4$ dex. After this cut, the plotted sample contains 17 galaxies (5 local dwarfs + 12 other dwarfs), excluding the highest-uncertainty cases from the Gannon/LVDB side (e.g. DF17, DFX1, DGSAT-I, and ESO349-031).

The ages and metallicities of MW GCs used in the right panel of Figure~\ref{fig1} are taken from \citetmet{VandenBerg:2013aa}.

Stellar abundances of MW GCs used in Figure~\ref{fig3} are from the SDSS Apache Point Observatory Galaxy Evolution Experiment (APOGEE) in the data release 17 \citemet{Abdurrouf:2022aa}. We use the allStarLite catalog with the same quality cuts as in \citetmet{Belokurov:2022aa} and we use the corrected [O/Fe] values as in \citetm{Ji:2026aa}. We only include red giant stars ($1.5<\log(g)<3$, $T_\mathrm{eff}<5300$\,K). We require errors on [O/Fe] and [N/Fe] to be less than 0.1\,dex. We use the globular cluster catalog of \citetmet{Vasiliev:2021aa} to identify member stars.

\clearpage

\bibliographymet{mainbib}

\section*{Acknowledgements}

The authors gratefully acknowledge the Gauss Centre for Supercomputing e.V. (\url{www.gauss-centre.eu}) for funding this project by providing computing time on the GCS Supercomputer SuperMUC-NG at Leibniz Supercomputing Centre (\url{www.lrz.de}), under project pn29we. WM thanks the Science and Technology Facilities Council (STFC) Center for Doctoral Training (CDT) in Data Intensive Science at the University of Cambridge (STFC grant number 2742968) for a PhD studentship. WM and ST acknowledge support by the Royal Society Research Grant G125142. RK acknowledges support of the Natural Sciences and Engineering Research Council of Canada (NSERC) through a Discovery Grant and a Discovery Launch Supplement (funding reference numbers RGPIN-2024-06222 and DGECR-2024-00144) and York University's Global Research Excellence Initiative. XS acknowledges the support from the National Aeronautics and Space Administration (NASA) theory grant JWST-AR-04814.

\newpage

\section*{Author information}

\subsection*{Affiliations}
\noindent
\hypertarget{inst:Kavli}$^{1}$Kavli Institute for Cosmology, University of Cambridge, Madingley Road, Cambridge CB3 0HA, UK
\\
\hypertarget{inst:Cav}$^{2}$Cavendish Laboratory, University of Cambridge, 19 JJ Thomson Avenue, Cambridge CB3 0HE, UK
\\
\hypertarget{inst:IoA}$^{3}$Institute of Astronomy, University of Cambridge, Madingley Road, Cambridge CB3 0HA, UK
\\
\hypertarget{inst:York}$^{4}$Department of Physics and Astronomy, York University, 4700 Keele Street, Toronto ON M3J 1P3, Canada
\\
\hypertarget{inst:Dallas}$^{5}$Department of Physics, The University of Texas at Dallas, Richardson, TX 75080, USA
\\
\hypertarget{inst:AIP}$^{6}$Leibniz-Institut f\"ur Astrophysik Potsdam, An der Sternwarte 16, Potsdam 14482, Germany
\\
\hypertarget{inst:IPMU}$^{7}$Kavli IPMU (WPI), The University of Tokyo, Kashiwa, Chiba 277-8583, Japan
\\
\hypertarget{inst:MIT}$^{8}$Department of Physics, Kavli Institute for Astrophysics and Space Research, Massachusetts Institute of Technology, Cambridge, MA 02139, USA
\\
\hypertarget{inst:CIERA}$^{9}$Center for Interdisciplinary Exploration and Research in Astrophysics (CIERA), Northwestern University, 1800 Sherman Avenue, Evanston, IL 60201, USA
\\
\hypertarget{inst:NMSU}$^{10}$Department of Astronomy, New Mexico State University, 1320 Frenger Mall, Las Cruces, NM 88003-8001, USA
\\
\hypertarget{inst:Penn}$^{11}$Department of Physics and Astronomy, University of Pennsylvania, 209 South 33rd Street, Philadelphia, PA 19104, USA
\\
\hypertarget{inst:IfA}$^{12}$Institute for Astronomy, University of Edinburgh, Blackford Hill, Edinburgh EH9 3HJ, UK
\\
\hypertarget{inst:PKU}$^{13}$Department of Astronomy, Peking University, Beijing 100871, China
\\
\hypertarget{inst:CfA}$^{14}$Center for Astrophysics $|$ Harvard $\&$ Smithsonian, 60 Garden Street, Cambridge, MA 02138, USA

\end{document}